# Slow dynamics of disordered zigzag chain molecules in layered LiVS$_2$ under electron irradiation

**Naoyuki Katayama[1], Keita Kojima[1,2], Tomoki Yamaguchi[3], Sosuke Hattori[1], Shinya Tamura[1], Koji Ohara[2], Shintaro Kobayashi[2], Koudai Sugimoto[4], Yukinori Ohta[3], Koh Saitoh[5], Hiroshi Sawa[1]**

**Abstract**

**Electronic instabilities in transition metal compounds often spontaneously form orbital molecules, which consist of orbital-coupled metal ions at low temperature. Recent local structural studies utilizing the pair distribution function revealed that preformed orbital molecules appear disordered even in the high-temperature paramagnetic phase. However, it is unclear whether preformed orbital molecules are dynamic or static. Here, we provide clear experimental evidence of the slow dynamics of disordered orbital molecules realized in the high-temperature paramagnetic phase of LiVS$_2$, which exhibits vanadium trimerization upon cooling below 314 K. Unexpectedly, the preformed orbital molecules appear as a disordered zigzag chain that fluctuate in both time and space under electron irradiation. Our findings should advance studies on unprecedented soft matter physics realized in an inorganic material due to disordered orbital molecules.**

[1]Department of Applied Physics, Nagoya University, Furo-cho, Chikusa-ku, Nagoya, Aichi 464-8603, Japan,
[2]Diffraction and Scattering Division, Center for Synchrotron Radiation, Japan Synchrotron Radiation Research Institute, 1-1-1 Kouto, Sayo, Hyogo 679-5198, Japan,
[3]Department of Physics, Chiba University, 1-33 Yayoi-cho, Inage-ku, Chiba 263-8522, Japan,
[4]Department of Physics, Keio University, 3-14-1 Hiyoshi, Kohoku-ku, Yokohama, Kanagawa 223-8522, Japan,
[5]Advanced Measurement Technology Center, Institute of Materials and Systems for Sustainability, Nagoya University, Furo-cho, Chikusa-ku, Nagoya, Aichi 464-8603, Japan

**Introduction**

Transition metal compounds with multiple electron degrees of freedom frequently form self-organized molecules in low temperature, called orbital molecules[1-16]. Examples include octamers in $CuIr_2S_4$[1] and trimers in $LiVO_2$ and $LiVS_2$[5-6]. Recent structural studies clarified that the orbital molecules at low temperature can persist in the high-temperature paramagnetic phase as a disordered form in some substances[9-16]. Persistent orbital molecules contradict the general belief that the regular lattice should recover in the high-temperature paramagnetic phase, as speculated based on the average structure obtained from conventional diffraction experiments[1-10]. Because the long-range ordering of the arrangements of orbital molecules are restored upon cooling in these substances, it is expected that the persistent orbital molecules realize unprecedented lattice dynamics. Kimber *et al.* predicted that a classical analogue of the short-range Resonating Valence Bond state will be realized in the high-temperature phase of $Li_2RuO_3$ as the existing dimer patterns resonate due to thermal fluctuations[13]. They explained that slower dynamics should appear compared to the characteristic timescale of their x-ray measurements, although these dynamics have yet to be experimentally clarified. By contrast, Browne *et al.* performed a quasi-elastic neutron scattering experiment up to 1100 K in $GaV_2O_4$ with disordered trimer and tetramer pairs in anticipation of the appearance of fast dynamics. However, they concluded that clusters are well-defined and statically disordered even at 1100 K as they were unable to detect dynamic behaviours faster than $\sim 1\times10^{-11}$ s [12].

Here, we present comprehensive structural studies of $LiVS_2$, which exhibits a paramagnetic metal to nonmagnetic insulator transition at 314 K[5]. A recent structural study clarified that vanadium trimer molecules form in the low-temperature nonmagnetic phase[6]. The vanadium trimer molecules disappear above 314 K, and unprecedented zigzag chain molecules consisting of dimers emerge with a finite correlation length. Cooling increases the correlation length, and sharp superstructure peaks grow in powder diffraction patterns below approximately 350 K.

Our annular dark-field scanning transmission electron microscope (ADF-STEM) experiment clearly shows that these existing zigzag chain molecules slowly fluctuate in both time and space on the order of seconds. The high-temperature phase with slow dynamics of some existing zigzag chain patterns is categorised as a plastic crystal phase, which is a mesophase between a crystal and liquid. The atoms retain the original position in the time average but the orientation of the zigzag chain fluctuates.

**Experimental Results**

**Physical properties of $LiVS_2$**

Figure 1**a** displays the physical properties of LiVS$_2$. It exhibits a metal to nonmagnetic insulator transition around 314 K. As a previous study has already clarified[5], the high-temperature paramagnetic susceptibility increases upon heating, which is reminiscent of pseudogap behavior found in underdoped high-$T_c$ cuprates. Although the present data are consistent with those reported previously, the present entropy change of $\Delta S$ = 7.99 J/mol K at 314 K is much larger than $\Delta S$ = 6.6 J/mol K in the previous study[5]. This difference may be due to the improved sample quality in this study. For checking the quality of the present sample, we performed the Inductively Coupled Plasma-Atomic Emission Spectrometry (ICP-AES) experiment, and confirmed that the Li content, *x*, is 0.97(2) in Li$_x$VS$_2$. A subtle amount of off-stoichiometry does not have a significant effect on the physical properties, as mentioned above. Note that no discontinuous behaviors appear in these physical property data above 314 K for the discussion later.

**Rietveld analysis**

Figure 1**b** shows the synchrotron x-ray diffraction patterns as functions of temperature. Below 314 K, sharp superstructure peaks emerge around $Q$ ~ 3.2-3.3 Å$^{-1}$. Recently, the low-temperature crystal structure was solved using a Rietveld method by assuming a trigonal space group $P31m$. It was clarified that the vanadium trimers in a spin singlet state spontaneously form (Figure 1**c**)[6].

Upon heating above 314 K, the superstructure peaks associated with vanadium trimerization completely disappear in the powder diffraction data and unprecedented additional superstructure peaks emerge (Figure 1**b**, arrows), which has never been reported. The additional peaks gradually weaken upon further heating and eventually disappear at approximately 350 K. For the higher temperature data, refinement can be successfully performed by assuming a trigonal space group $P\bar{3}m1$ with a regular triangular lattice, while the refinement can be successfully performed for powder diffraction data obtained at 320 K by assuming a monoclinic space group $Pm$ with vanadium zigzag chain molecules (Figure 1**d**). Strangely, the distinct superstructure peaks clearly show the monoclinic displacement of the consisting atoms, the monoclinic lattice strain from the triangular lattice is negligible. Both the temperature dependent lattice parameters and the details of the Rietveld refinement are summarized in the Supplemental Information. Figure 1**e-g** display the high-resolution x-ray diffraction patterns at 320 K with the indices. The emergences of the superstructure peaks with $l \neq 0$ in the indices indicate that the zigzag chain appears as three-dimensional ordering.

**PDF analysis**

Fig. 1: Temperature dependences of the physical properties of LiVS$_2$ with structure deformations.

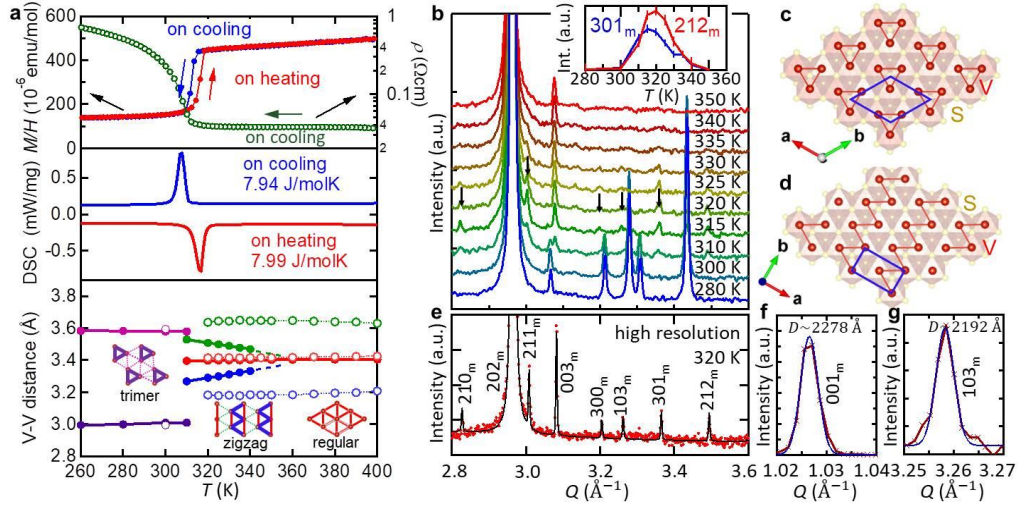

**a**, Temperature dependences of the magnetic susceptibility and resistivity (top), DSC data (middle), and adjacent V–V distances (bottom). Open circles are derived by analysing the reduced PDF data in the range of $2 \leq r (\text{Å}) \leq 10$, while the closed circles are from the Rietveld analysis of the powder diffraction data. Insets schematically depict the averaged structure obtained from Rietveld analysis in each temperature range. **b**, Temperature dependences of powder x-ray diffraction data. Arrows indicate peaks that appear just above the trimer transition temperature of 314 K, which weaken upon further heating and disappear around 350 K. Detailed results of Rietveld analysis for each temperature regions were summarized in Supplemental Information. Inset shows the temperature dependences on superstructure peak intensity. **c**, Schematic of vanadium trimers realized below 314 K. **d**, Schematic of vanadium zigzag chain appearing with a finite correlation length in the high-temperature paramagnetic phase. **e**, High resolution x-ray diffraction patterns at 320 K with indices of monoclinic phase. **f-g**, Enlarged views of a fundamental 001 peak (**f**) and a superstructure 103 peak (**g**) with the correlation length estimated using the Scherrer equation. Instrumental resolution is not taken into account. These data are obtained from the high-resolution data.

Pair Distribution Function (PDF) analysis was performed to evaluate whether the preformed orbital molecules appear in a disordered form in the high-temperature paramagnetic phase. Figure 2**a** displays the reduced PDF data in the range of $2 \leq r (\text{Å}) \leq 10$ obtained at 325 K, where the averaged $Pm$ structure is realized, although the monoclinic lattice strain from the triangular lattice is negligibly small. At first, the fit assuming the average $P31m$ model with the vanadium trimers is poor ($R_w = 18.6\%$) because the

Fig. 2: Reduced Pair Distribution Function (PDF) analysis of LiVS$_2$ in the high-temperature paramagnetic phase.

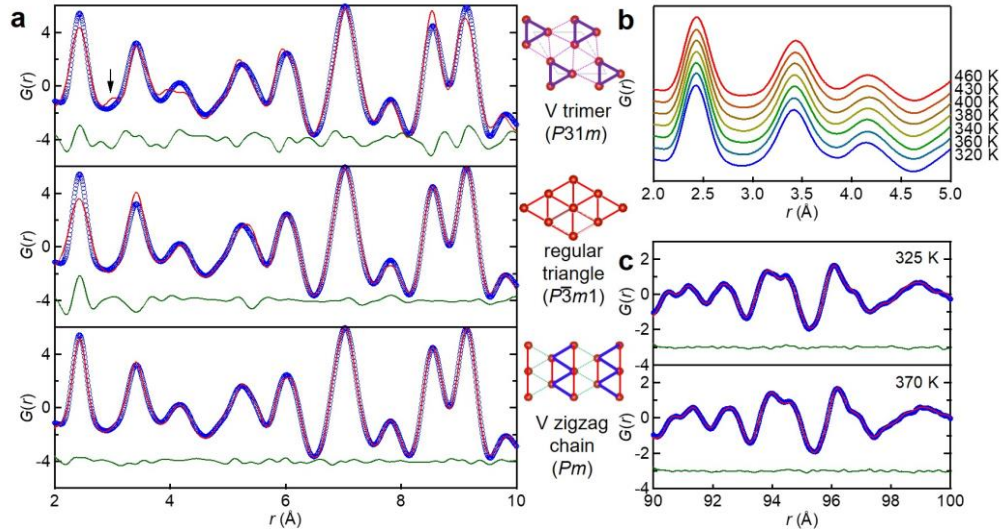

**a**, Fitting (red) and residual (green) for reduced PDF data (blue) obtained at 325 K. Average $P31m$ model with V trimers (top) generates unrealized peak around 3.0 Å, resulting in a poor fit. The $P\bar{3}m1$ model with regular triangular lattice cannot be fitted well to the intensity data (middle). Best fit is obtained by assuming the $Pm$ structure with a zigzag chain (bottom). Adjacent V–V distances can be separated into three types, which are drawn in blue, red, and green in the $Pm$ structure. Blue is the shortest and forms a zigzag chain. Initial structural model is obtained from the Rietveld refinement of the powder diffraction experiment at the corresponding temperature. **b**, Temperature dependences of PDF data in the high-temperature paramagnetic phase. Increasing the temperature up to 460 K does not cause a remarkable change in the PDF patterns, indicating that the zigzag chain model can be applied up to at least 460 K. **c**, PDF analysis in the range of $90 \leq r\,(\text{Å}) \leq 100$ by assuming the $Pm$ structure both for 325 K and 370 K data. The reliable factors are $R_w = 6.04\%$ (325 K) and 5.29% (370 K), respectively.

simulation generates an unrealized peak at $r \sim 3.0$ Å, which corresponds to the inner-trimer V–V distance. This clearly indicates that disordered vanadium trimers are absent in the high-temperature paramagnetic phase. Secondly, the fit assuming the average $P\bar{3}m1$ model with the regular vanadium lattice is also poor ($R_w = 14.6\%$). The peak at $r \sim 3.4$ Å, which corresponds to adjacent V–V and S–S distances, is much broader than the peak at $r \sim 2.5$ Å, which consists of the adjacent V–S distance. We cannot successfully fit both peaks by assuming the average $P\bar{3}m1$ model even if we assume anomalous thermal parameters for V and S ions. If the average $P\bar{3}m1$ model is realized, the peak at $r \sim 3.4$ Å should be stronger and sharper. Therefore, we can also exclude the regular triangle from the candidates. The experiment confirmed that the monoclinic $Pm$ crystal structure with a vanadium zigzag chain gives the best fit ($R_w = 5.37\%$), consisting with the Rietveld analysis results. Unexpectedly, sharply different from the powder diffraction patterns shown in Figure 1**b**, where the superstructure peaks disappear at approximately 350 K, the peak profiles in the range of $2 \leq r\,(\text{Å}) \leq 10$ are maintained without any significant changes upon heating at least up to 460 K (Figure 2**b**). Hence, the average $Pm$ model can still be fitted well to the reduced PDF patterns even at 460 K. The V–V distances obtained from the refined structures using Rietveld and the PDF analysis in the range of $2 \leq r\,(\text{Å}) \leq 10$ exhibit significant differences (Figure 1**a**, bottom). We can successfully explain it in terms of the correlation length of the $Pm$ monoclinic domain, as described below.

The displacement of V obtained from Rietveld refinement is always smaller than that obtained from PDF analysis and decreases with increasing temperatures. Thus, the correlation length of the monoclinic domains with zigzag chain molecules has a finite value above 314 K and decreases upon heating. It should be noted that the superstructure peaks are as sharp as the fundamental peaks at 320 K, as seen in Figures 1**f** and 1**g**. This means that the width of the superstructure peaks is limited by the instrumental width. From Scherrer's equation, we can roughly estimate the correlation length of monoclinic domains to be longer than ~2200 Å at 320 K. Although the superstructure peaks disappear above approximately 350 K in the powder diffraction data, the monoclinic domains persist up to at least 460 K (Figure 2**b**, reduced PDF data). Because the superstructure peaks disappear in the continuous process where the correlation length shortens upon heating, anomalies do not appear in the physical property data around 350 K (Figure 1**a**). As shown in Figure 2**c**, the overall appearances of PDF data collected at 325 K and 370 K are quite similar between them even in the range of $90 \leq r\,(\text{Å}) \leq 100$, and refinement is successfully performed by assuming the $Pm$ crystal structure with a zigzag chain, indicating the long correlation length over 90 Å is maintained even at 370 K. Further discussion about the PDF analysis at high $r$ regions is available at the Supplementary Information of this article.

**STEM experiment**

The strong temperature dependence on correlation length expects us that the zigzag chain molecules have dynamic properties. To clarify whether the preformed zigzag chain molecules are dynamic or static, we observed a time series of high-resolution annular dark-field scanning transmission electron microscope (ADF-STEM) images of LiVS$_2$ at room temperature. In the thick part of the piece away from the edge, such as in the lower left portion of the inset of Figure 3**a**, the diffraction pattern derived from the $P31m$ phase having

Fig. 3: ADF-STEM

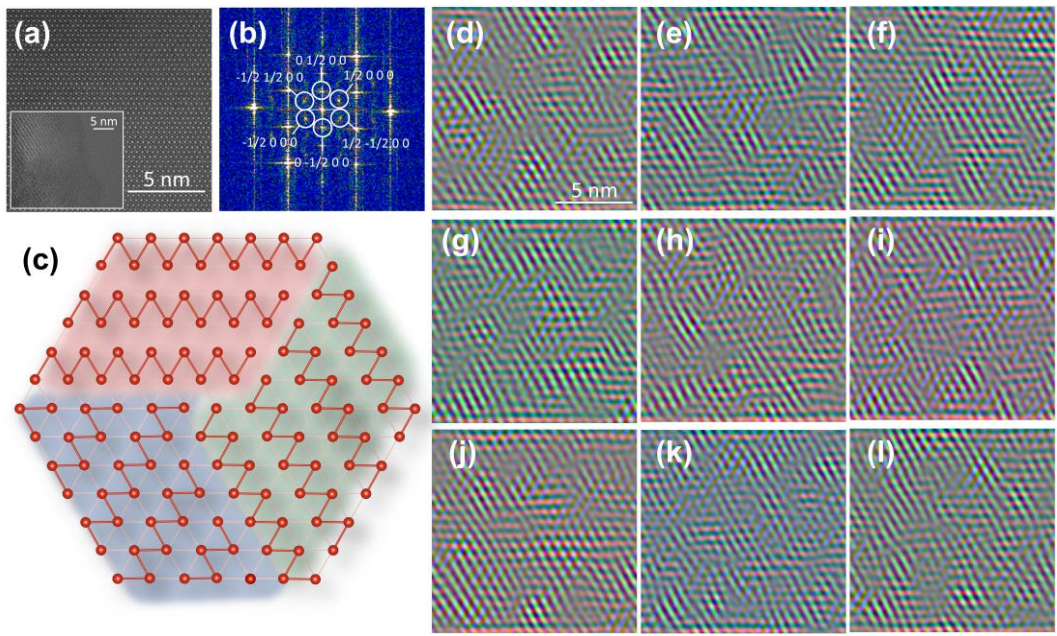

**a**, High-resolution ADF-STEM image on the edge region observed along the [0001] incidence at room temperature. Inset shows a real space image of a small LiVS$_2$ piece observed at room temperature. **b**, Fourier transform pattern of the ADF-STEM image. **c**, Schematic of existing zigzag chain patterns. Potential patterns are resonating possibly due to thermal fluctuations. Please see the main text for more details. **d–l**, Time series of Fourier-masked ADF-STEM images obtained at intervals of almost one second. The corresponding movie file is available (Supplemental Information).

a trimer was clearly observed. The obtained diffraction pattern is similar with that previously reported[5]. On the other hand, unexpectedly, a diffraction pattern derived from the monoclinic $Pm$ phase with a zigzag chain was observed despite the measurement at room temperature in the region near the edge of the sample. The ADF-STEM experimental results on this edge region will be presented below.

Figure 3**a** (1024 × 1024 pixels) shows a high-resolution ADF-STEM image observed by [0001] incidence on the edge of the specimen. Image areas of approximately 130 Å × 130 Å were acquired with a pixel dwell time of 1 microsecond, which is a time interval of approximately 1 second. STEM images were taken continuously, and time series were acquired at intervals of almost 1 second. The bright and dark dots correspond to V and S atom columns projected along the [0001] orientation, respectively. The V and S atoms form a two-dimensional triangular lattice. The Fourier transform pattern of the ADF-STEM image shows superlattice spots at positions $\pm 1/2\ 0\ 0\ 0$, $0\ \pm 1/2\ 0\ 0$, and $\mp 1/2\ \pm 1/2\ 0\ 0$ for high temperature trigonal unit cell (Figure 3**b**). This indicates the presence of three

monoclinic domains, which are rotated by 120 degrees from each other (Figure 3c). Figures 3**d–l** show a time series of Fourier-masked ADF-STEM images. The inverse FFT is performed using the superlattice spots of $\pm 1/2\ 0\ 0\ 0$, $0\ \pm 1/2\ 0\ 0$, and $\mp 1/2\ \pm 1/2\ 0\ 0$, and the real space distribution of each domain is colored by red, green and blue. The intensity of the three types of lattice fringes is modulated in different ways to form a unique domain structure. Note that these images show the structures projected in the beam direction, where domain patterns appearing in several VS$_2$ layers are superimposed. Remarkably, the modulation and domain patterns change in every frame of the time series of STEM images, demonstrating that the zigzag chain molecules fluctuate on a timescale of seconds (The corresponding movie file is available). The slow dynamics on a timescale of seconds clearly indicates that the present state is caused by thermal fluctuations.

Our STEM experimental results raise the question of why the $Pm$ phase, which should occur above 314 K, occurs in the edge region at room temperature. There are several possible scenarios. First, it is possible that the sample temperature rose due to electron beam irradiation and exceeded the phase transition temperature of 314 K. But this is probably not happening. When a carbon thin film is irradiated with a 5 nA beam with a diameter of 1 nm, the temperature rise is reported to be 1.4 K[17]. In this experiment, the beam diameter was about 0.1 nm and the electron beam current was 0.1 nA. This indicates that the sample has a much lower electron dose per unit time. Furthermore, this scenario cannot be explained that the $P31m$ phase with vanadium trimers was also observed in the region away from the edge under the same experimental conditions. The second scenario concerns Li being unexpectedly missing at the edge of the sample. In the Li$_x$VS$_2$ system, it has been reported that the trimer transition temperature decreases as the Li content $x$ decreases from 1.0, and the trimer transition temperature falls below room temperature when $x$ = 0.8. Note that our x-ray diffraction experimental results of the Li deficient samples are consistent with this tendency. The superlattice peaks survive at 300 K for $x$ = 0.91(1), while they disappear at ~ 250 K for $x$ = 0.78(1) (Supplemental Information). When the transition temperature is significantly lowered from room temperature due to Li deficiency, the correlation length of the $Pm$ phase at room temperature may be sharply shortened, and multiple domains may be superimposed in this ADF-STEM image at room temperature. Another possibility within this scenario is that the Li at the edges is strongly diffused at room temperature due to the thermal and/or electron irradiation effects. A third possible scenario is that the phase transition temperature can change with sample thickness, as is often the case with some transition metal dichalcogenides[18,19]. For example, in 1T-TaS$_2$, the low temperature Commensurate Charge Density Wave (CCDW) phase is suppressed when the sample thickness is below the threshold of ~ 40 nm[19]. Layer thicknesses at the edge of current samples are expected to be up to a few nanometers since it is not possible to perform ADF-STEM experiments on thicker samples. A fourth possible scenario is that the anomalous charge-fluctuating phase accompanied by lattice dynamics is induced by electron irradiation. A reference phenomenon has been observed in $Ba_3NaRu_2O_9$, which contains the $Ru_2O_9$ dimer, where the charge order melts under photoexcitation[20]. Further experimental studies should be required to clarify which scenario is realized in the future.

**Theoretical considerations**

Fig. 4: Calculated electronic structures of $LiVS_2$.

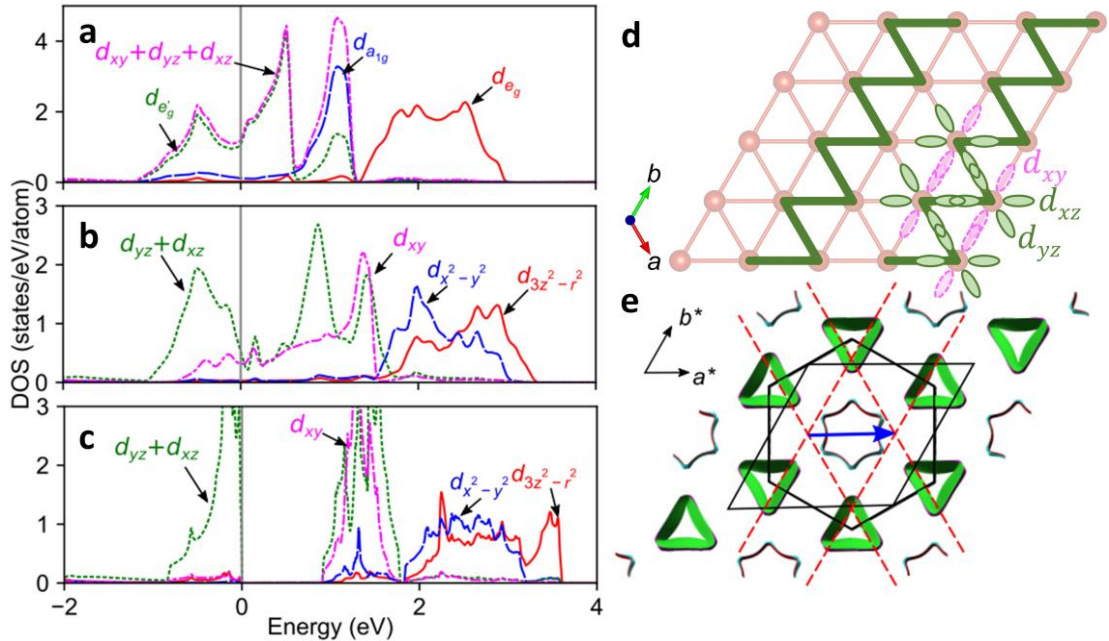

**a**, Orbital-decomposed partial densities of states (PDOS) in the paramagnetic metallic phase with the average $P\bar{3}m1$ structure; the $d_{xy} + d_{yz} + d_{xz}$ component is separated into the $e'_g$ and $a_{1g}$ components in the trigonal crystal field. **b**, Orbital-decomposed PDOS calculated for the local $Pm$ structure. **c**, Orbital-decomposed PDOS calculated for the trimer structure. **d**, Orbital-assisted multiple dimerizations consisting of zigzag chain patterns. **e**, Fermi surfaces calculated using the GGA+U scheme at $U = 4$ eV.

After identifying the unprecedented dynamics of disordered zigzag chain molecules appearing in the high-temperature paramagnetic phase of $LiVS_2$, we considered the underlying physics generating such dynamics of disordered zigzag chain molecules. Figure

4**a** shows the band calculation result based on the parameters obtained from the Rietveld analysis of 360 K data by assuming the trigonal space group $P\bar{3}m1$. The triply degenerate $t_{2g}$ orbitals, which consist of $xy$, $yz$, and $xz$ orbitals, are oriented toward the neighbouring vanadium ions and are inherently separated into lower doubly degenerate $e_g'$ orbitals and a higher nondegenerate $a_{1g}$ orbital due to the trigonal splitting (Figure 4**d**).

The corresponding first principles calculation result for a $P\bar{3}m1$ structure with $U = 4$ eV suggests a Fermi surface with the possible common nesting vector of $Q = a^*/2$ (Figure 4**e**), which can cause multiple dimerizations of the lattice along the $a$- and $(a+b)$-direction. As shown in Figure 4**d**, this leads to the zigzag chain pattern. Our band calculation result obtained using the parameters of the local $Pm$ structure at $360$ K clearly shows that the $d_{yz} + d_{xz}$ component, which is due to the doubly degenerate $e_g'$ bands in the high-temperature $P\bar{3}m1$ phase, is separated into bonding and antibonding bands. Electrons fill the bonding bands (Figure 4**b**). The Fermi surface survives due to the remnant $d_{xy}$ component even when the calculation assumes the $Pm$ structure consisting of a metallic nature in high-temperature paramagnetic phase. We naively insist that the persisting Fermi surface in the $Pm$ phase weakens the band energy stabilisation associated with the zigzag chain formation as it may be related to the emergence of the observed slow dynamics. When vanadium trimers are formed in the lower temperature region, a large band gap consisting of the insulative behavior in an electrical resistivity experiment (Figure 4**c**). We speculate that the transition at 314 K induces a large band energy at the expense of lattice energy.

**Discussion**

As discussed above, the short-range order of the zigzag chains appearing in the high temperature phase of LiVS$_2$ can be originating from the orbitally assisted CDW instability. C. Rovira and M.-H. Whangbo discuss that both zigzag chains and trimers are explained on the basis of the concept of both local chemical bonds and hidden Fermi surface nesting[21]: When $t_{2g}$ orbital is triply degenerated and the proportionate band filling of ($d_{xy}$, $d_{yz}$, $d_{zx}$) = (2/3, 2/3, 2/3) is realized, the trimerization appears as a result of the Fermi surface nesting. On the other hand, when the disproportionate band-filling of ($d_{xy}$, $d_{yz}$, $d_{zx}$) = (1,1,0) is realized, the zigzag chain pattern appears as the ground state. Based on this argument, one may consider that the phase transition at 314 K in LiVS$_2$ originates from two energetically competing CDW states. However, the situation is not so simple. First, in the high temperature phase of LiVS$_2$, the triply degeneracy $t_{2g}$ orbital is not maintained, and the nesting instability that causes zigzag chain order is inherently present. In other words, the zigzag chain order can be formed without causing band disproportionation. Second, there is no band instability that stabilizes trimerization. This indicates that the phase transition at 314 K is accompanied by another ingredient which changes band structure, such as the band Jahn-Teller transition.

Both trimers and zigzag chains are composed of vanadium ions with two bonds, attributed by the $V^{3+}$ $d^2$ electronic state of $LiVS_2$. Since the zigzag chain has a finite correlation length in the high temperature phase, vanadium with only one bond must be present at the end of the zigzag chain. This may cause pseudogap-like behavior of magnetic susceptibility temperature dependence. When bond formation occurs, spin singlets are formed inside the bond, which makes it essentially nonmagnetic. However, since the vanadium at the end of the chain can form only one bond, the extra electrons should contribute to the paramagnetism. As the correlation length of the zigzag chain becomes shorter upon heating, the number of vanadium at the edge increases, should be leading to the enhancement of Pauli paramagnetism upon heating, as shown in Figure 1**a**. Previous studies clarified that the enhancement of paramnagnetic susceptibility continues up to at least 700 K, possibly indicating that the short-range order survives even 700 K.

Although our theoretical calculations indicate that the Fermi surface instability may be critical to form the zigzag chain patterns, the origin of the unprecedented dynamics is unclear. Considering that the disordered zigzag chain consists of the orbital-coupled metal ions, the unconventional interaction between orbital and phonon should be an ingredient for generating the unusual lattice dynamics. However, the slow dynamics on a second timescale indicate that the fluctuation itself is not of an electronic origin. Phenomenologically, it is reminiscent of the physics of soft materials such as a polymer rotation. The high-temperature phase of $LiVS_2$ with slow dynamics of zigzag chain patterns does not belong to a crystal where both the centre of gravity and the orientation of the atoms/molecules should be maintained. Considering that the atoms retain the original position in a time average but the zigzag chain orientation fluctuates over time, we categorise it as a plastic crystal, which is a mesophase[22,23].

Conventionally, plastic crystals are realized in weakly interacting molecules or ions, where the consisting molecules/ions are thermally rotating at a fixed position. The dynamics of conventional plastic crystals are usually studied via NMR techniques. Considering that NMR covers the kHz-MHz order, we can estimate that the zigzag chain dynamics realized in the present $LiVS_2$ should be some order of magnitude slower than the rotation dynamics observed in conventional plastic crystals. We speculate that this is due to the strong interaction among neighbouring atoms, which depend on the complex network structure of inorganic materials. Our findings should provide a new platform for investigating soft matter physics in inorganic materials as well as expand the fundamental understanding of systems with preformed orbital molecules at high temperature.

In summary, we first observed the slow dynamics of disordered orbital molecules appearing in the high-temperature paramagnetic phase of LiVS$_2$. The unconventional coupling between orbital and phonon should be an ingredient for generating the unprecedented dynamics, surely impact the studies of conventional orbital physics, such as nematic state in iron selenides[24-26]. Motivated by this work, we expect that further explorations of dynamics targeting similar systems, such as Li$_2$RuO$_3$ and AlV$_2$O$_4$, should be accelerated, leading to the novel research fields of disordered orbital molecules.

## Materials and Methods

### Sample growth and preparation

Powder samples of LiVS$_2$ were prepared by a soft-chemical method and a subsequent solid-state reaction. Li-deficient Li$_{-0.75}$VS$_2$ was synthesised initially by a reaction with an appropriate amount of Li$_2$S, V, and S in an Ar-filled quartz tube at $700\,°C$ for $3$ days. Note that the obtained precursors include small single crystalline samples with the order of $\mu$m. The products were put in 0.2 M *n*-BuLi hexane solution for $2$ days to attain the maximum Li content[27]. The Li content, $x$, was confirmed to be 0.97(2) in Li$_x$VS2 from the Inductively Coupled Plasma-Atomic Emission Spectrometry (ICP-AES) using Hitachi SPCTRO ARCOS, equipped at the Institute of Solid State Physics (ISSP), Japan. The samples were crashed well to make powder sample to perform the x-ray diffraction experiments. In a STEM experiment, the sample of the same batch was diffused in anhydrous hexane and roughly ground for about 30 seconds. The supernatant was placed on a sample holder and used for measurement.

### Physical property measurements

Differential scanning calorimetry (DSC) was conducted using a DSC 204 F1 Phoenix (Netzsch). The magnetic susceptibility was measured by a SQUID magnetometer (Quantum Design). The electrical resistivity was measured by the four-probe method. The powder samples were sintered at a low temperature of 300 $°C$ because the inserted Li ions were partially deintercalated at higher temperatures. Experiments were performed in an Ar atmosphere.

### Powder diffraction experiments

To investigate the average structure, synchrotron powder x-ray diffraction experiments were performed with the incident x-ray energy of $E = 19\,\text{keV}$ in BL5S2 beamline at Aichi Synchrotron, Japan, and with the incident x-ray energy of $E = 30\,\text{keV}$ in BL02B2 beamline at SPring-8, Japan. While figure 1**b** was constructed by the data obtained in BL5S2, the data of figure 1**e-f** was collected in BL02B2 beamline. A two-dimensional semiconductor detector, PILATUS 100K, was used for high-resolution measurements and high-speed data collection in the experiments performed in BL5S2 beamline, while the six MYTHEN detectors were used in BL02B2 beamline. RIETAN-FP software[28] was employed for the Rietveld analysis, while VESTA was used for graphical purposes[29]. High-energy synchrotron x-ray diffraction experiments with $E = 61$ keV were performed at BL04B2 at SPring-8, Japan, to study the local structure from PDF analysis. Hybrid point detectors of Ge and CdTe were employed for data collection. The reduced PDF $G(r)$ was obtained via the conventional Fourier transform of the collected data[30]. The PDFgui package[31] was used to analyse $G(r)$. Capillaries containing samples and empty capillaries were measured under similar optical conditions, and the measurement results of the empty capillaries were subtracted from the data as the background. Polarization correction, Compton correction, and absorption correction were applied to obtain the reduced PDF data. For making reduced PDF data, the diffraction data with the $Q$ range of $0.2 \leq Q(\text{Å}^{-1}) \leq 25.7$ with the $\Delta Q = 0.011$ were used. Delta 1 and delta 2 options were not utilized in PDFGui software.

**Scanning transmission electron microscopy**

A time series of high-resolution scanning transmission electron microscope (STEM) images were obtained with a spherical aberration–corrected electron microscope equipped with a cold-type field emission gun, JEOL JEM-ARM 200F, operating at 200 kV. The convergent semi-angle of the incident beam was set to 20 mrad. The inner and outer semi-angles of the annular detector were set to 65 mrad and 265 mrad, respectively. A series of the STEM images with 1024 × 1024 pixels were acquired with a pixel dwell time of 1 microsecond. Although the specimen did not show irradiation damage in the STEM observation time series, the electron beam might increase the temperature in the illuminated areas, inducing the thermal fluctuation observed in the study.

The fast Fourier transform (FFT) patterns and Fourier-masked images were obtained by computer software, Gatan Digital Micrograph. The diameters of the masks used to select the superlattice reflections were set to 1 nm$^{-1}$.

**Computational details**

We employed the WIEN2k code[32] based on the full-potential linearised augmented plane-wave method for our density functional theory (DFT) calculations. The calculated results obtained in the generalised gradient approximation (GGA) for electron correlations were presented with the exchange-correlation potential of Ref. 33. To improve the description of electron correlations in V $3d$ orbitals, we used the rotationally invariant version of the GGA+$U$ method with the double-counting correction in the fully localised limit[34,35]. The main text presents the results obtained at $U = 4$ eV. In the self-consistent calculations, we used 16 × 16 × 7, 10 × 10 × 18, and 13 × 13 × 11 $\boldsymbol{k}$-points in the irreducible wedges of the Brillouin zones for the $P\bar{3}m1$, $Pm$, and $P31m$ phases, respectively. Muffin-tin radii ($R_{MT}$) of 2.15 (Li), 2.35 (V), and 2.03 (S) Bohr were used assuming a plane-wave cutoff of $K_{max} = 7.00/R_{MT}$. We used VESTA[29] and XCrySDen[36] for graphical purposes.

**Acknowledgements**

The authors acknowledge Prof. M. Cuoco, Prof. K. Nishikawa and Prof. M. Moriya for valuable discussions. N.K. and K.K. acknowledge Dr. Ishii for technical support for determining Li amount using Hitachi SPCTRO ARCOS. The work leading to these results has received funding from the Grant in Aid for Scientific Research (No. JP20H02604, No. JP17K05530, No. JP19K14644, No. JP17K17793, JP20H01849 and JP19J10805), Keio University Academic Development Funds for Individual Research, the Thermal and Electric Energy Technology Inc. Foundation, and Daiko Foundation. This work was carried out under the Visiting Researcher's Program of the Institute for Solid State Physics, the University of Tokyo, and the Collaborative Research Projects of Laboratory for Materials and Structures, Institute of Innovative Research, Tokyo Institute of Technology. The synchrotron powder x-ray diffraction experiments for Rietveld analysis were conducted at the BL5S2 of Aichi Synchrotron Radiation Center, Aichi Science and Technology Foundation, Aichi, Japan (Proposals No. 201704027, No. 201801026, No. 201802042, No. 201803046, No. 201804016, No. 201806026, No. 201901018 and No. 201902056), and at the BL02B2 beamline of SPring-8, Hyogo, Japan (Proposals No. 2019B1073 and No. 2018B1157). The high energy synchrotron powder x-ray diffraction experiments for PDF analysis were conducted at the BL04B2 of SPring-8, Hyogo, Japan (Proposals No. 2018B1128 and No. 2018B1145).

**Author information**

**Contributions**

N.K. conceived the original idea. N.K. and K.Sa. planned and designed the experiments.

K.K. and S.T. fabricated the samples under the supervision of N.K.. N.K. performed the transport measurements and magnetic susceptibility measurements. K.K. and S.T. performed the DSC measurements. N.K., K.K., S.T. and H.S. performed the synchrotron powder diffraction experiments and performed the analysis of the relevant data. N.K., K.K., S.K. and K.O. performed the high energy powder diffraction experiments and K.K. performed the PDF analysis under the supervision of K.O. S.H. and K.Sa. performed the scanning transmission electron microscopy experiments and analyzed the relevant data. T.Y., K.Su. and Y.O. performed the theoretical calculation based on the structural parameters obtained from powder diffraction experiments. N.K. wrote the paper with K.Sa. and Y.O. All the authors discussed the results and commented on the manuscript.

**Corresponding author**

Correspondence to Naoyuki Katayama.

**Ethics declarations**

**Competing interests**

The authors declare no competing interests.

# Supplementary Information for
# Slow dynamics of disordered zigzag chain molecules in layered LiVS$_2$ under electron irradiation

Naoyuki Katayama[1], Keita Kojima[1,2], Tomoki Yamaguchi[3], Sosuke Hattori[1], Shinya Tamura[1], Koji Ohara[2], Shintaro Kobayashi[2], Koudai Sugimoto[4], Yukinori Ohta[3], Koh Saitoh[5], Hiroshi Sawa[1]

**Supplementary Note 1 : Refinement details**

In this section, we display the Rietveld and PDF refinement results at various temperatures with the structural parameters obtained therein. The cif files obtained from the Rietveld analysis of the data collected at 400 K, 320 K and 200 K are available as supplemental data of this article, and through the CCDC database with the deposition numbers of 1957344-1957346.

**Supplementary Note 2 : Lattice parameters and isotropic atomic displacement parameters**

Supplementary Figure 1 displays the temperature dependences on lattice parameters collected from the Rietveld analysis of the data. First, we explain how we determined the crystal structure of $Pm$ phase.

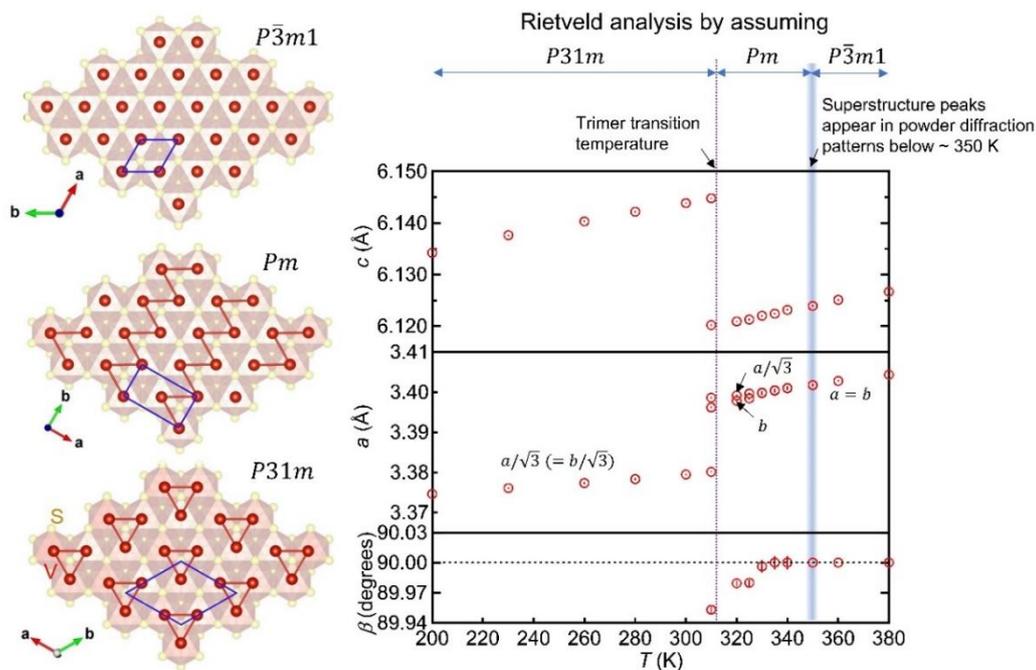

Supplementary Figure 1 : Temperature dependences of lattice parameters attained from Rietveld analysis.

From the Bragg peak positions, we can easily find the primitive monoclinic unit cell with $b_{\text{monoclinic}} = b_{\text{trigonal}} \times \sqrt{3}$ is realized at 320 K, as shown in Supplementary Figure 1 left, although the monoclinic distortion can be hardly found. Furthermore, we can also find extinction rule does not exist at 320 K diffraction patterns, resulting that the possible space group can be limited to $P2$, $Pm$ or $P2/m$. After some Rietveld analysis trials, it was clarified that no significant atomic displacement due to the two-fold symmetry occurred, but only the atomic displacement derived from the mirror symmetry appeared. Therefore, we can uniquely determine the space group to be $Pm$ at 320 K.

While a big jump appears corresponding to the trimer transition at around 314 K, the lattice parameters change smoothly across the transition from $P\bar{3}m1$ to $Pm$ at around 350 K. This is consistent with the fact that the transition happens on the increasing process of the correlation length. Correspondingly, the $\beta$ angle retains almost 90 degrees despite of the averaged monoclinic structure. Note that the $\beta$ angle (= 90 degree) of the $P\bar{3}m1$ phase is defined by assuming a monoclinic unit cell.

Supplementary Figure 2 shows isotropic atomic displacement parameters ($B$) of sulfur and vanadium, obtained from the Rietveld experiment of powder diffraction data. For refinements, a $P31m$ structure (trimer structure) is assumed for low temperature data below 314 K, while a $P\bar{3}m1$ structure (regular triangle structure) is assumed for all high temperature data above 314 K. Although the superstructure peaks appear between 314 K and 350 K, trigonal space group of $P\bar{3}m1$ is employed for refinement by ignoring the superstructure peaks.

Although both $B_{\text{sulfur}}$ and $B_{\text{vanadium}}$ show jump at 314 K, the jump of $B_{\text{vanadium}}$ is larger than $B_{\text{sulfur}}$. The $B_{\text{vanadium}}$ at high temperature reaches to anomalously large value of 1.2-1.4. This expects us that the disorder originates from vanadium displacement. $B_{\text{vanadium}}$ value is almost constant throughout the high temperature phase. This is consistent with the fact that the local V-V distance, which can be obtained from the PDF data, is constant in the temperature range measured.

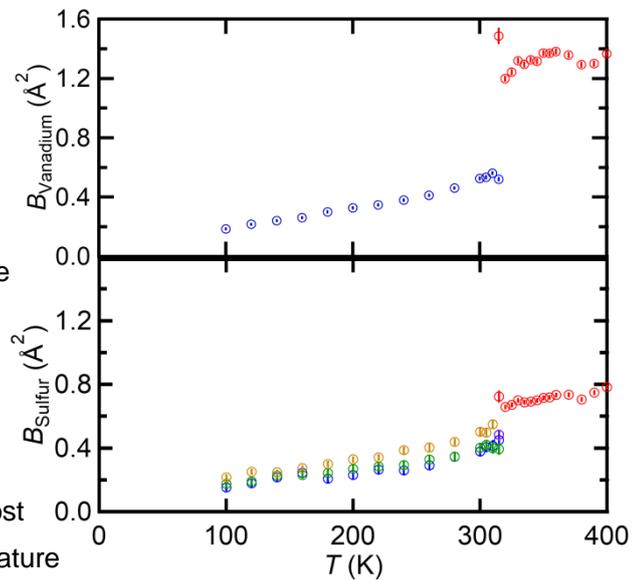

Supplementary Figure 2. Atomic displacement parameters for sulfur and vanadium. Note that there are three sulfur sites in $P31m$ phase.

**Supplementary Note 3 : Structural analysis results of the powder diffraction data (trimer phase)**

Supplementary Figure 3 displays x-ray diffraction data obtained at SPring-8 BL02B2 beamline and the Rietveld refinement results. The data are collected at 300 K, where the trimer structure appears in average structure. Although the superstructure peaks, which appear at 1/3 1/3 0 and its related positions for high temperature trigonal cell, are fitted well both in intensity and width, no special techniques are employed for refinement, such as individual peak shape fitting. Although a small amount of Li deficient phase (2.8%) was included as an impurity phase, we successfully co-refined with a main phase of $LiVS_2$. The lattice parameters for Li deficient phase was derived from the previous report (D.W. Murphy *et al.*, Inorg. Chem. **16** (1977) 3027.). The insets show the expanded data on the logarithmic scale.

The cif file at 200 K is available through the CCDC database with the number of 1957345.

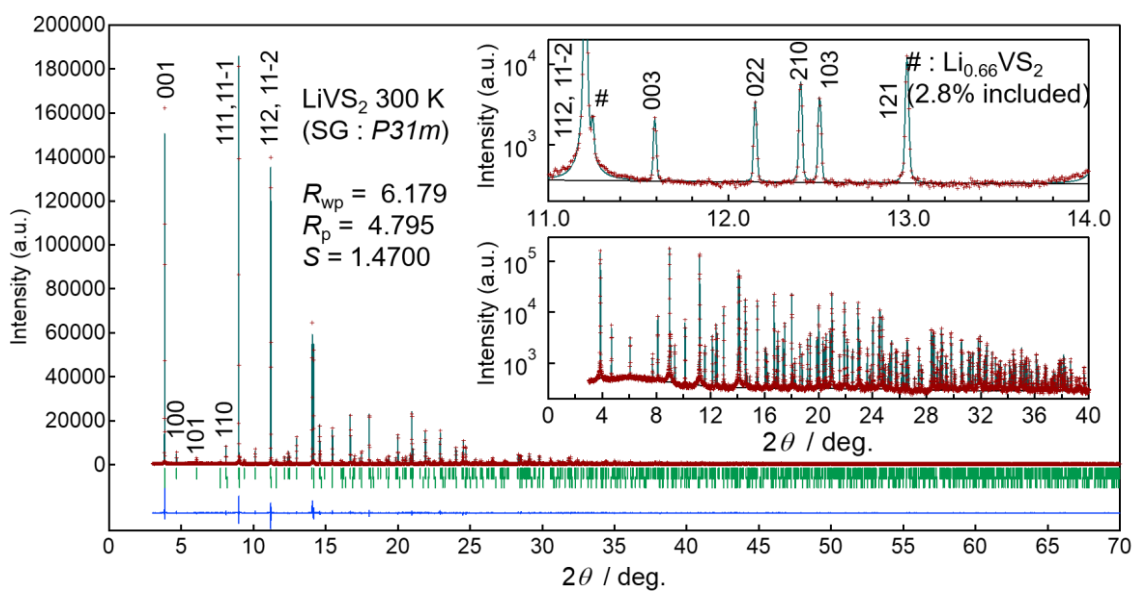

Supplementary Figure 3 : Rietveld analysis result of powder diffraction patterns obtained at 300 K. This is data from a batch sample different from the sample used for PDF analysis, and contains a small amount of impurities. Due to the small amount of impurities, the temperature factors and atomic positions are not refined for the second phase. Upper inset shows the expanded data on the logarithmic scale. Lower inset shows overall diffraction patterns on the logarithmic scale.

Supplementary Figure 4 indicates the correlation length estimated from the peak width using Scherrer equation for fundamental and superstructure peaks. We cannot find the significant broadening of the superstructure peaks. The peak widths are almost consistent among them.

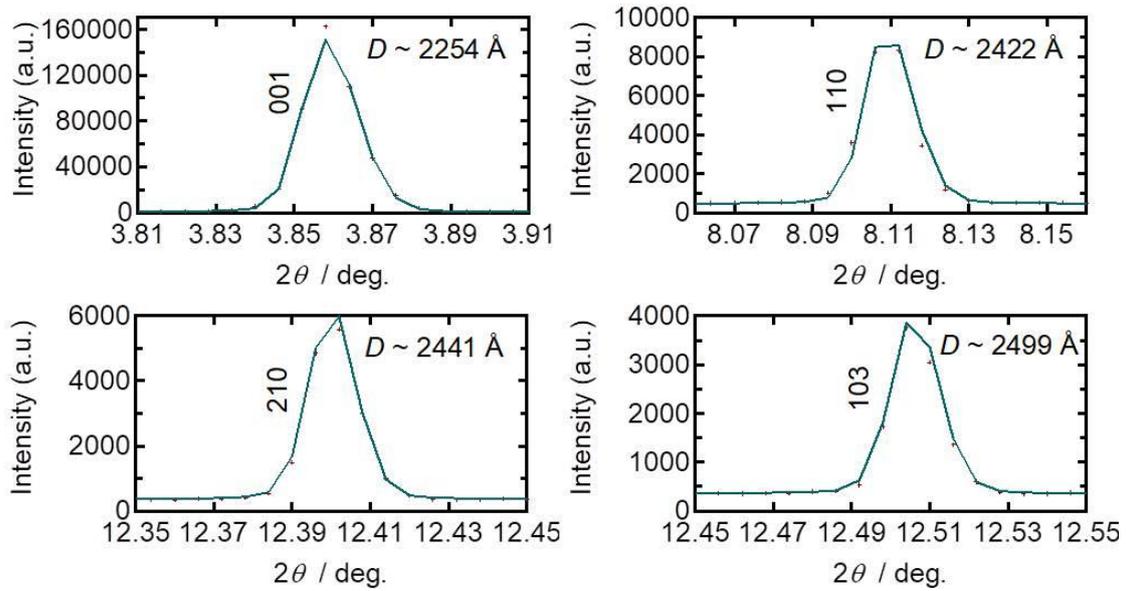

Supplementary Figure 4. Enlarged peaks for 300 K (P31m phase) data with a Rietveld fitting. 001 and 110 peaks are fundamental peaks, while 210 and 103 peaks are superstructure peaks.

Supplementary Figure 5 presents PDF analysis performed by assuming some possible structures. Reliable factors obtained by assuming structures of $P31m$ with trimers (top), $P\bar{3}m1$ with regular lattice (middle) and $Pm$ with zigzag chains (bottom) are $R_w = 4.58\%$, 21.8%, 14.6%, respectively. The peak shown by an arrow at $r\,(\text{Å}) \sim 3.00$ was fitted only when $P31m$ structure was assumed.

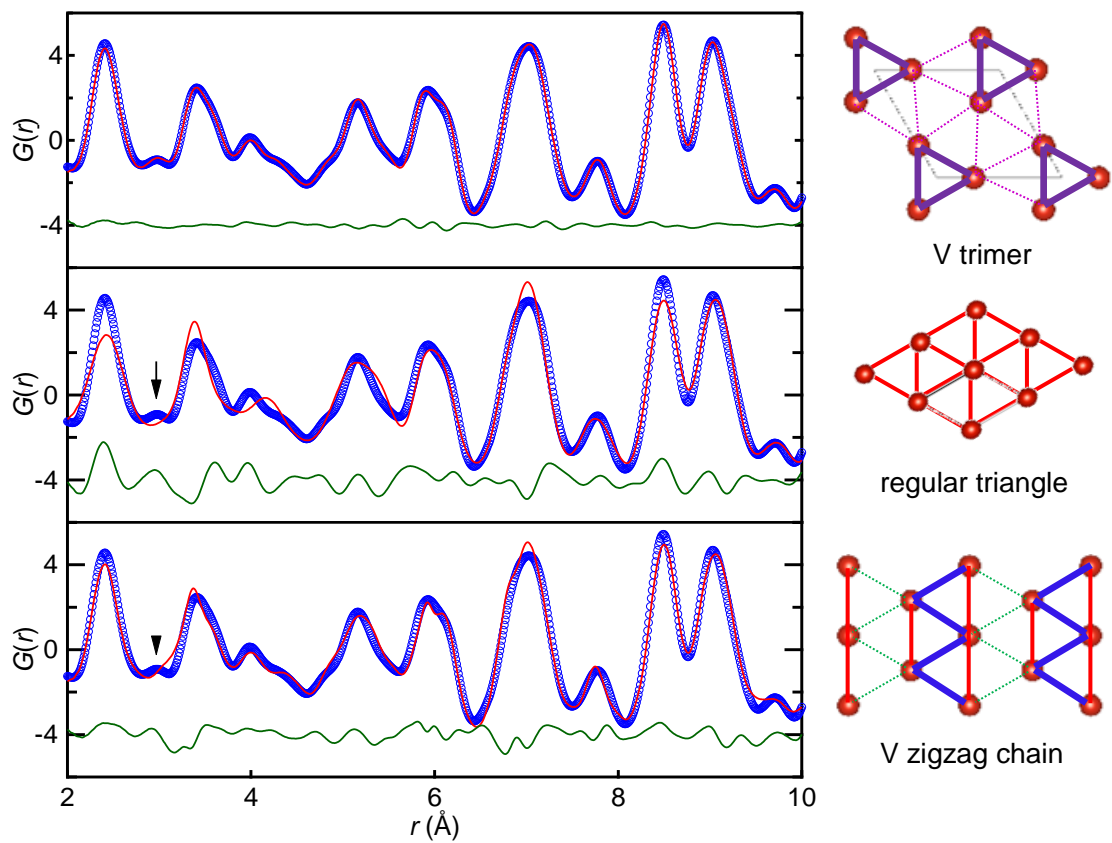

Supplementary Figure 5 : PDF analysis by assuming some possible structures performed at 300 K.

**Supplementary Note 4 : Structural analysis results of the powder diffraction data (zigzag phase)**

Supplementary Figure 6 displays x-ray diffraction data obtained at SPring-8 BL02B2 beamline and the Rietveld refinement results. X-ray diffraction data obtained at SPring-8 BL02B2 beamline and the Rietveld refinement results. The data are collected at 320 K, where the zigzag structure appears in average structure. Although the superstructure peaks are fitted well both in intensity and width, no special techniques are employed for refinement, such as individual peak shape fitting. Although a small amount of Li deficient phase (1.7%) was included as an impurity phase, we successfully co-refined with a main phase of $LiVS_2$. The lattice parameters for Li deficient phase was refined from the initial parameters of 300 K data. The insets show the expanded data on the logarithmic scale. In the Rietveld analysis, we fixed one vanadium site to be (0, 0, 0). This is because we need to fix x and z on one site when the space group $Pm$ is assumed, which is consisting of three atomic sites: 2c (x,y,z), 1b (x, 1/2, z) and 1a (x,0,z).

PDF analysis results at the almost corresponding temperature of 325 K is found in the main text of this paper. The corresponding cif file is available through the CCDC database with the deposition number of 1957346.

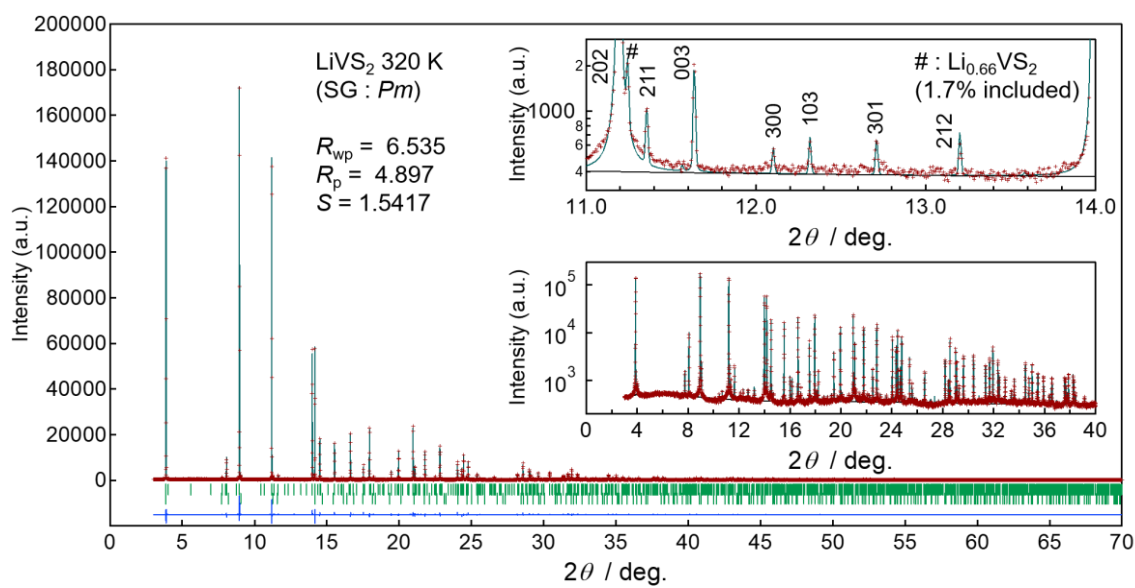

Supplementary Figure 6 : Rietveld analysis result of powder diffraction patterns obtained at 320 K. This is data from a batch sample different from the sample used for PDF analysis, and contains a small amount of impurities. Due to the small amount of impurities, the temperature factors and atomic positions are not refined for the second phase. Upper inset shows the expanded data on the logarithmic scale. Lower inset shows overall diffraction patterns on the logarithmic scale.

Supplementary Figure 7 indicates the correlation length estimated from the peak width using Scherrer equation for fundamental and superstructure peaks. The correlation length of the zigzag chain is longer than ~ 2200 Å because the peak width is apparently limited by the instrumental width. We cannot find the significant broadening of the peaks. The peak widths are almost consistent among them. The PDF patterns at 320 K are presented with the fitting in the main text of this article.

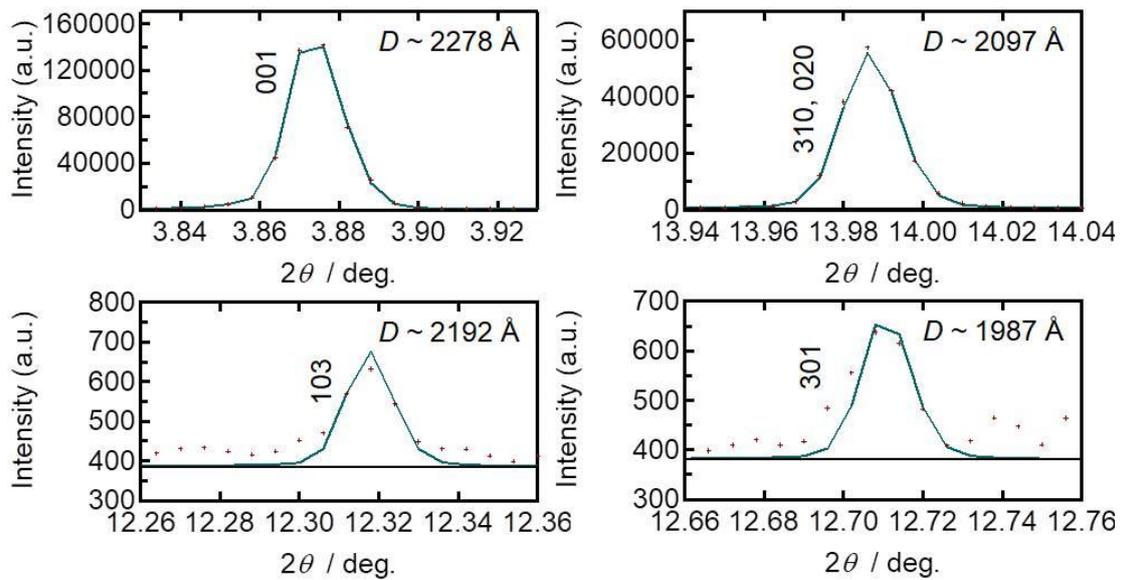

Supplementary Figure 7. Enlarged peaks for 320 K (Pm phase) data with a Rietveld fitting. 001 and 310(020) peaks are fundamental peaks, while 103 and 301 peaks are superstructure peaks. 310 and 020 peaks emerge at the close positions.

**Supplementary Note 5 : Structural analysis results of the powder diffraction data (regular phase)**

Supplementary Figure 8 displays x-ray diffraction data obtained at SPring-8 BL02B2 beamline and the Rietveld refinement results. The data are collected at 400 K, where the regular structure appears in average structure. Although a small amount of Li deficient phase (0.9%) was included as an impurity phase, we successfully co-refined with a main phase of $LiVS_2$. The lattice parameters for Li deficient phase was refined from the initial parameters of 320 K data. The insets show the expanded data on the logarithmic scale.

The corresponding cif file is available through the CCDC database with the deposition number of 1957344.

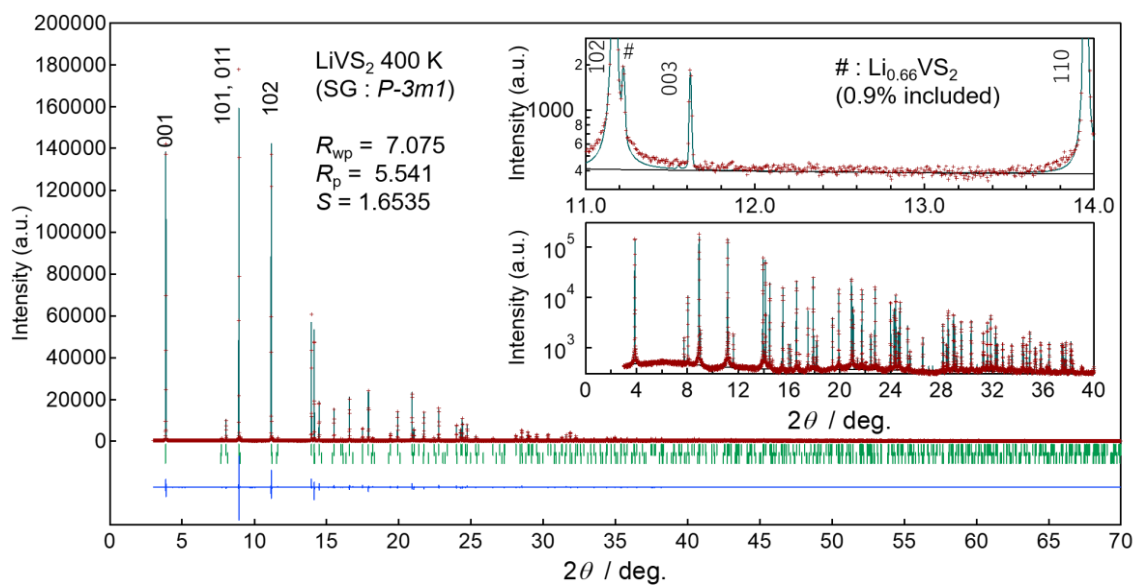

Supplementary Figure 8. Rietveld analysis result of powder diffraction patterns obtained at 400 K. This is data from a batch sample different from the sample used for PDF analysis, and contains a small amount of impurities. Due to the small amount of impurities, the temperature factors and atomic positions are not refined for the second phase. Upper inset shows the expanded data on the logarithmic scale. Lower inset shows overall diffraction patterns on the logarithmic scale.

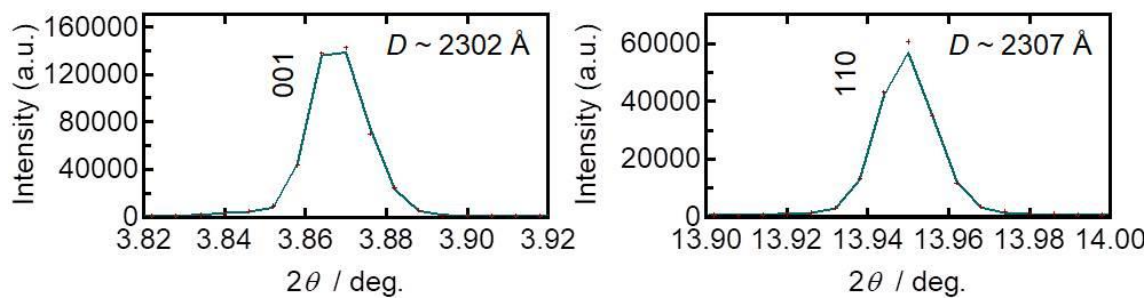

Supplementary Figure 9. Enlarged peaks for 400 K (P-3m1 phase) data with a Rietveld fitting. HWHM are almost consistent between these peaks.

Supplementary Figure 10 presents PDF analysis performed by assuming some possible structures. Reliable factors obtained by assuming structures of $P\bar{3}m1$ with regular lattice (top) and $Pm$ with zigzag chains (bottom) are $R_w = 12.0\%$ and $5.32\%$, respectively. In the refinement by assuming $P\bar{3}m1$ structure, the peak at around $r\,(\text{Å}) \sim 3.43$ must be comparable or higher in intensity compared with the peak at around $r\,(\text{Å}) \sim 2.43$, which is contradictory to the experimentally obtained data.

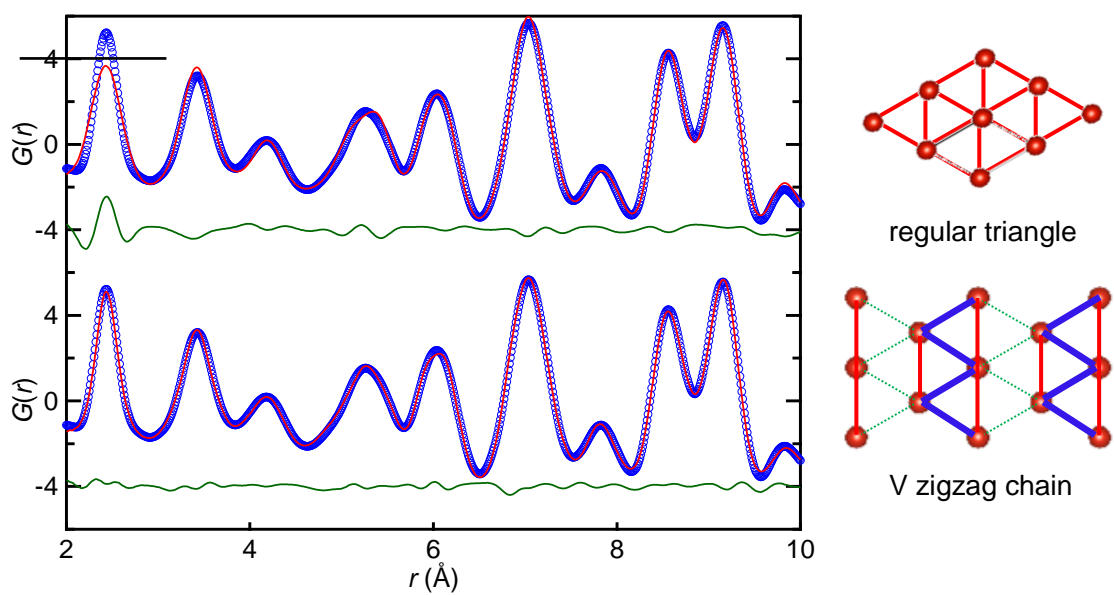

Supplementary Figure 10 : PDF analysis by assuming some possible structures performed at 400 K.

**Supplementary Note 6 : Q[S(Q)-1] vs Q plot**

Here we show the Q dependences on Q[S(Q)-1] data at various temperature ranges. The data was used to obtain the reduced G(r) data for PDF analysis. This plot allows one to see if the diffraction signal is anomalously damped due to short range disorder, and also to see what short-range order is left at high-Q. As shown in Supplementary Figure 11, we can clearly find the Q[S(Q)-1] data at 325 K and 400 K show similar Q dependencies, which support that the similar short-range ordering is realized at 325 K and 400 K. These data are quite different from 300 K data.

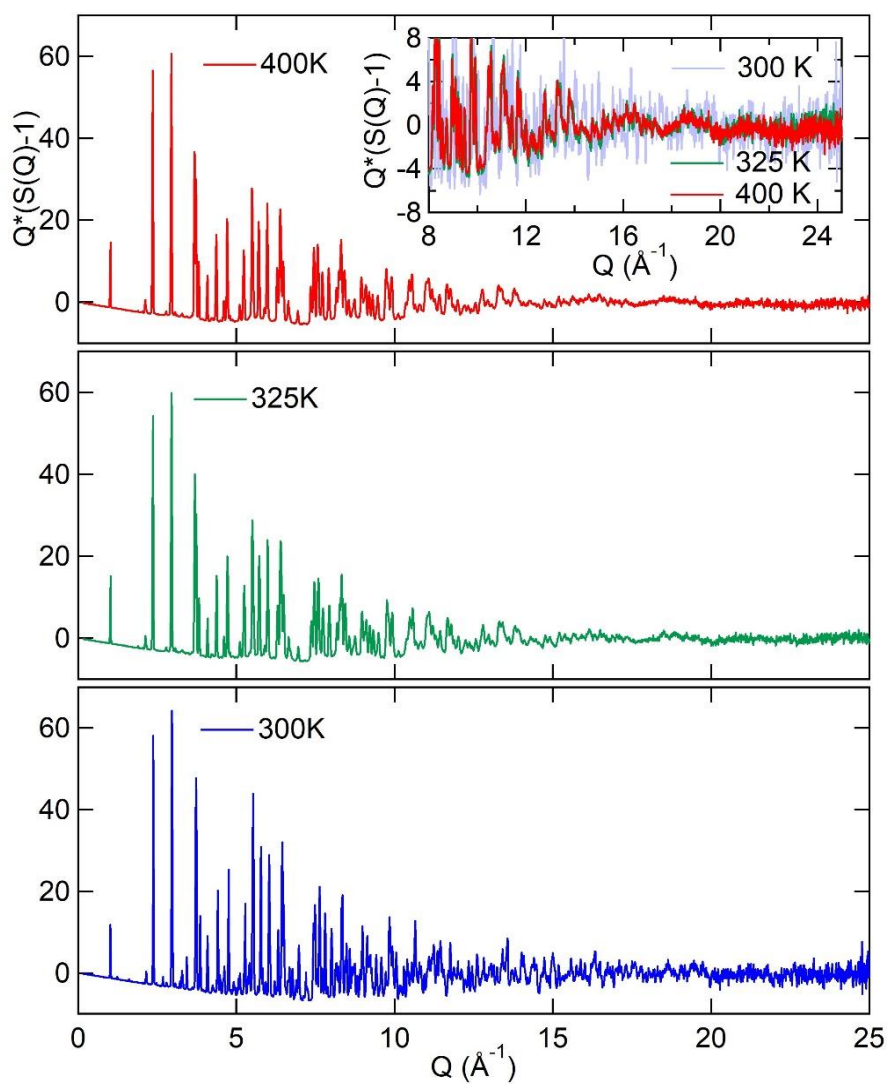

Supplementary Figure 11 : Q dependences on Q[S(Q)-1] data obtained at 300 K, 325 K and 400 K.

**Supplementary Note 7 : Estimation of the correlation length.**

Just above the $T_{trimer}$, the correlation length is long enough to generate sharp superstructure peaks in powder diffraction experiments, indicating zigzag chain formation. The widths of the superstructure peaks are almost comparable to those of Bragg peaks, as shown in Figure 1**e-g** in the main text. Although the accurate estimation of correlation length based on the width of superstructure peaks cannot be performed because the width of superstructure peaks are almost limited by the instrumental resolution, it should be noted that the correlation length roughly estimated by applying Scherrer's equation to 103 peak of $Pm$ structure at 320 K, which appears at around $Q \sim 3.26$ as shown in Figure 2**g**, reaches to $\sim 2200$ Å.

Supplementary Figure 12 displays the reliable factor, $R_w$, obtained by refining the data obtained at 325 K and 370 K by assuming the $Pm$ structure for various $r$ ranges. We can clearly find that $R_w$ is almost consistent between the data at 325 K and 370 K. This again indicates that the correlation length is longer than $\sim 100$ Å even at 370 K. We cannot perform the PDF analysis for higher $r$ regions because the reduced PDF $G(r)$ becomes too noisy for performing the reliable PDF analysis.

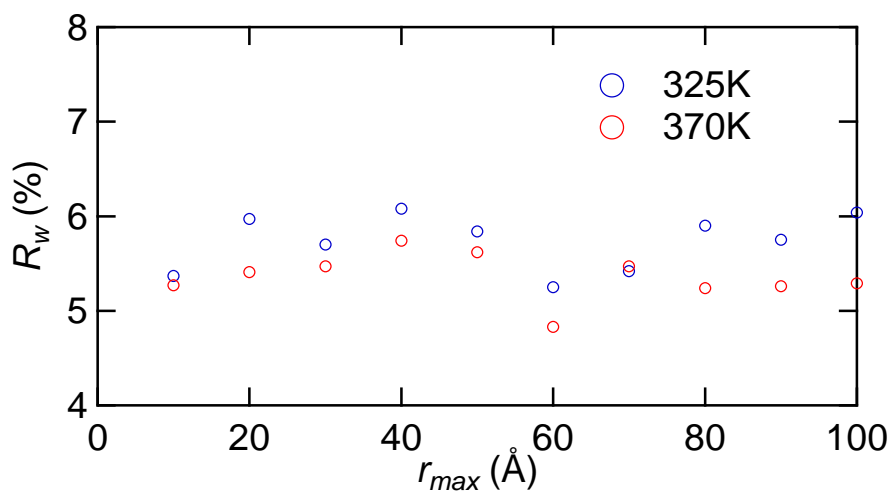

Supplementary Figure 12 : The reliable factor, $R_w$, as a function of $r_{max}$. The refinement was performed in the range of $r_{max} - 10 \leq r \leq r_{max}$.

**Supplementary Note 8 : Synchrotron powder diffraction patterns of Li deficient samples.**

Here, we present the synchrotron x-ray diffraction patterns of Li deficient samples. Li deficient samples were prepared by the soft chemical technique using $I_2$ acetonitrile solution[S1]. We can find that the superstructure peaks of $P\bar{3}m1$ phase are robust even at 300 K in $x$ = 0.91(1), while the peaks disappear at around 250 K in $x$ = 0.78(1). Note that the Li content was determined using the lattice parameter $a$ at 400 K. It is known that the lattice parameter $a$ linearly changes depending on the composition from $x$ = 0 to $x$ = 1 at 400 K[S2].

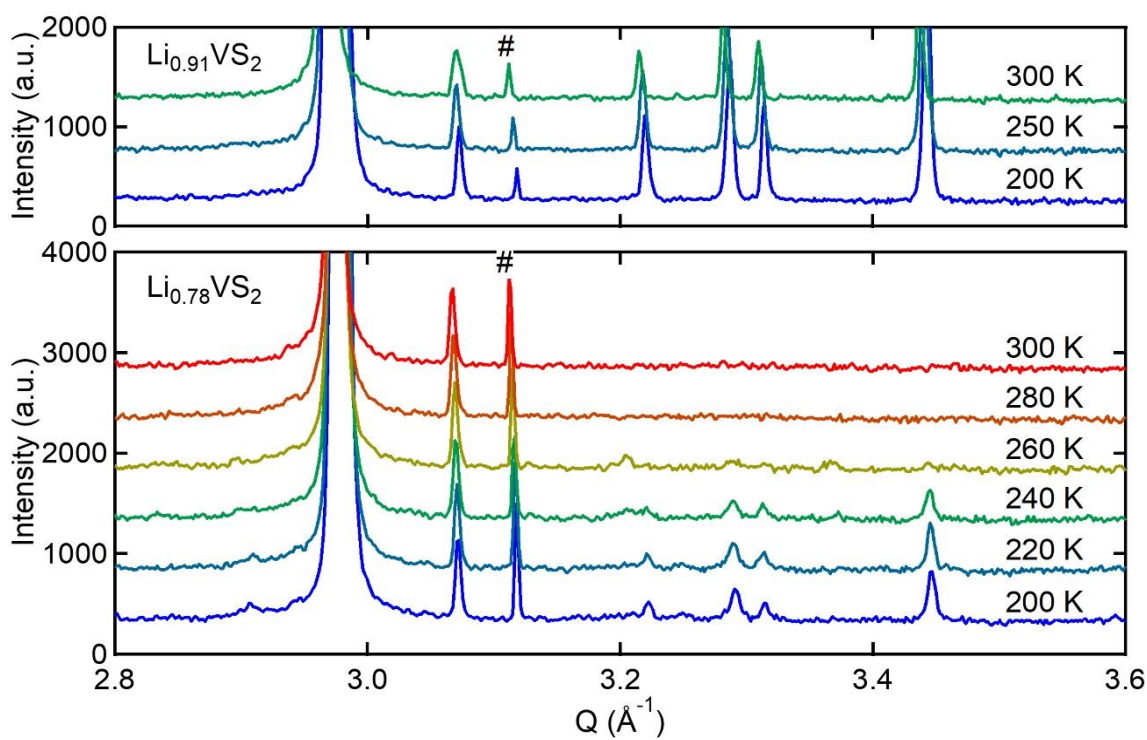

Supplementary Figure 13 : Synchrotron X-ray diffraction patterns for $x$ = 0.91(1) and $x$ = 0.78(1) in Li$_x$VS$_2$. Data was attained in BL02B2 beamline equipped at SPring-8, Japan. The incident x-ray energy of $E$ = 30 keV was used. # indicates the impurity of Li$_2$S.

**Supplementary References**

**Supplementary Note 9 : Time-series of ADF-STEM experiment**

A movie file of the time series images (STEM_dynamics_movie.gif) is available as an attached file of supplemental information. Movie legend is shown below.

> Supplementary Movie I : Time series images of Fourier-masked ADF-STEM images. Data obtained at intervals of almost one second were connected to form the movie. Blue, red and green regions indicate the monoclinic domains with different zigzag orientation.

**Supplementary Note 10 : Phonon dispersion in the high temperature phase investigated using the first-principles calculation**

To directly investigate the instability of the crystal structure of the LiVS$_2$ with a trigonal space group $P\bar{3}m1$, we performed the first-principles calculation of phonon dispersions, using the QuantumESPRESSO package[27,28] with the revised Perdew-Burke-Ernzerhof generalized gradient approximation[29] and the projector augmented-wave pseudopotential by Kresse and Joubert[30,31]. The plane-wave cut-off energy was set to 150 Ry, and the *k*-point mesh on the 12 × 12 × 6 Monkhorst–Pack grid[32] was used. The Methfessel-Paxton scheme[33] with a smearing width of 0.01 Ry was applied since the electronic structure is metallic. The cell structure was fully relaxed before calculating the dynamical matrix, and the interatomic force constants were obtained via the density functional perturbation theory (DFPT)[34] on the 4 × 4 × 2 grid.

The phonon dispersion of LiVS$_2$ with a trigonal space group $P\bar{3}m1$ is illustrated in Supplementary Figure 14. The acoustic phonon branch shows a negative-value frequency, implying that the crystal structure used in the calculation is unstable. The most negative value of the phonon frequency appears at the M point, and the corresponding wave vector coincides with the nesting vector calculated in the main text. We also confirm that the associated eigenvectors at the M point give the zigzag-chain-type atomic displacements (Supplemental data, M.gif)[S3]. Therefore, the analysis by DFPT also supports the fact that LiVS$_2$ with a trigonal space group $P\bar{3}m1$ has the instability to form the zigzag chain.

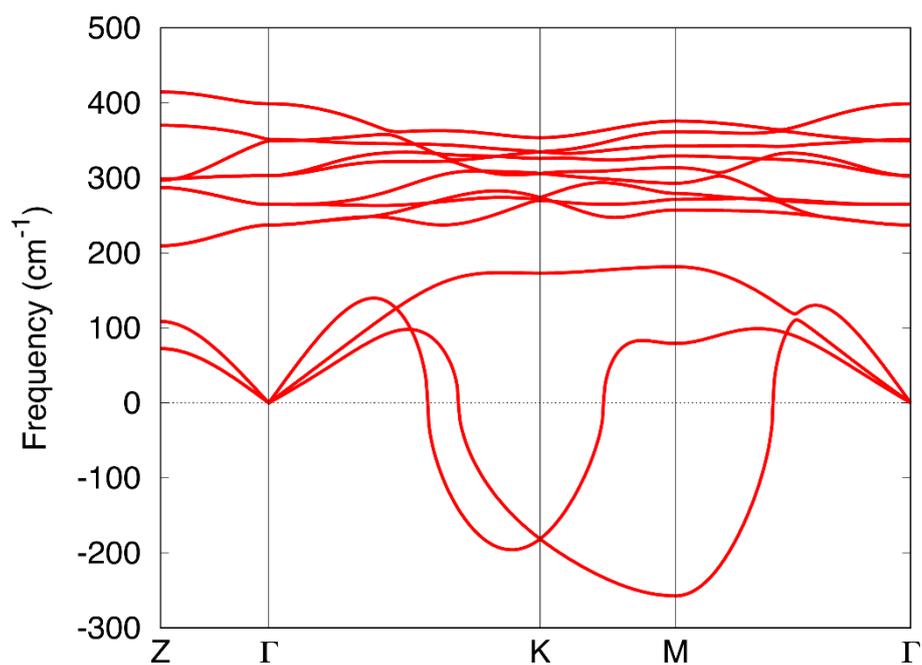

Supplementary Figure 14 : Calculated phonon dispersion of LiVS$_2$ with a trigonal space group $P\bar{3}m1$. The most negative value of the frequency appears at the M point.

**Supplementary Reference**

[S3]  http://henriquemiranda.github.io/phononwebsite/